\documentclass[twocolumn]{revtex4-2}
\usepackage{lineno}
\usepackage{graphicx}
\usepackage{dcolumn}
\usepackage{bm}
\usepackage{comment}
\usepackage{hyperref}
\usepackage{amsmath}
\usepackage{xcolor}

\begin{document}

\title{Observation of the centrality-dependent difference in directed flow between charged kaons 
and $K^{*0}$ resonances in Au+Au collisions at $\sqrt{s_{\mathrm{NN}}}$ = 14.6, 19.6 and 27 GeV}

\author{The STAR Collaboration}
\affiliation{}
\begin{abstract}
We present the measurement of rapidity-odd directed flow ($v_{1}^{\mathrm{odd}}$) for charged kaons, 
$\phi$ mesons, and $K^{*0}(\overline{K^{*0}})$  resonances in Au+Au collisions at $\sqrt{s_{\mathrm{NN}}}$ = 14.6, 19.6 and 27 GeV. 
This study includes the first measurement of the $K^{*0}$ resonance $v_{1}^{\mathrm{odd}}$ in heavy-ion 
collisions. Our measurement shows a centrality-dependent difference in directed flow between charged 
kaons and $K^{*0}$ resonances, which becomes more pronounced with increasing collision centrality. 
In contrast, the difference in directed flow between charged kaons and $\phi$ mesons remains nearly independent 
of centrality. Although anisotropic flow is thought to be developed in the early stages of the collision, for 
short-lived resonances such as the $K^{*}(892)$ that experience substantial hadronic re-scattering, 
it remains unclear to what extent the observed $v_{1}^{\mathrm{odd}}$ reflects genuine partonic collectivity in the early stage, and to 
what extent it is altered by late-stage hadronic interactions and/or reconstruction effects. The present measurement is 
crucial for disentangling these contributions and addressing this question. Existing 
hydrodynamic calculations that include a hadronic afterburner based on the UrQMD model indicate that 
an asymmetric modification of the $K^{*0}$ yield, due to rescattering, relative to the first-order 
event plane is required to reproduce the observed difference between $v_{1}^{\mathrm{odd}}$ for charged kaons and $K^{*0}$. 

\end{abstract}
\maketitle


%
\section{Introduction}
\label{S:1}

High-energy nuclear collisions provide an environment for studying strongly interacting QCD matter 
at extreme temperature and density. These collisions are expected to produce a deconfined state of 
matter consisting of quarks and gluons, known as the quark-gluon plasma (QGP)~\cite{STAR:2005gfr}. 
In a hydrodynamic picture, this medium begins in a pre-equilibrium stage and evolves into a thermalized QGP. As the system expands and cools, it transitions into the hadronic phase near the QCD critical temperature. Inelastic collisions cease at chemical 
freeze-out, fixing the primordial particle yields, while elastic interactions persist until kinetic 
freeze-out.~\cite{Becattini:1997rv,Braun-Munzinger:2001hwo,Xu:2001zj}

Short-lived resonances are sensitive probes of the hadronic phase in heavy-ion collisions, since their properties can be altered by interactions within the dense hadronic medium~\cite{Brown:1991kk,Markert:2008jc,Schaffner-Bielich:1999cux,Rapp:1999ej,STAR:2006vhb}. 
The measured yield of the $K^{*0}$ meson is particularly sensitive because its lifetime ($\sim$ 4 fm/c) is comparable 
to the duration of the hadronic medium~\cite{Rapp:2000gy}. It decays via 
$K^{*0}(\overline{K^{*0}}) \rightarrow K^{\pm}\pi^{\mp}$ with a branching ratio of 66.66\% ~\cite{ParticleDataGroup:2024cfk}. The decay daughters can rescatter, hindering $K^{*0}$ 
reconstruction, or  new $K^{*0}$ mesons can be regenerated after chemical freeze-out through 
pseudo-elastic interactions of pions and kaons produced in the 
medium~\cite{Bleicher:2003ij,Knospe:2015nva,Steinheimer:2017vju,Bleicher:2002dm}. 
The competition between these processes is reflected in the $K^{*0}/K^{\pm}$ ratio. A reduction in the $K^{*0}/K^{\pm}$ ratio with increasing event multiplicity indicates rescattering dominance, while enhancement implies significant 
regeneration. Measurements show reduced $K^{*0}/K^{\pm}$ ratios in heavy-ion collisions compared to small 
systems like $p+p$, consistent with increased hadronic rescattering effects in larger 
systems~\cite{STAR:2002npn,STAR:2022sir,STAR:2026kqj,ALICE:2021ptz,Sahoo:2023rko,Sahoo:2023dkv}. On the other hand, owing to its relatively long lifetime ($\sim$ 46 fm/$c$), the $\phi$ meson is largely unaffected by hadronic re-scattering. Therefore, comparisons between the $\phi$ and $K^{*0}$ mesons are often used to investigate the effects of hadronic re-scattering in the late stages of the collision evolution.

\textit{Flow} observables are characterized by the coefficients ($v_n$) in the Fourier expansion of azimuthal distributions of produced particles and they provide insight into the medium's evolution~\cite{Ollit:1992}. $v_n$ is defined by 
\begin{equation}
    v_{n} =\langle \rm{cos}\ \textit{n}(\phi-\psi) \rangle,
\end{equation}
where $\phi$ and $\psi$ denote the azimuthal angles of an outgoing particle and a reaction plane, respectively~\cite{art_pos}. The first harmonic coefficient is known as the directed flow 
($v_{1}$). 
Hydrodynamic and transport models show that $v_{1}$ is sensitive to early-time dynamics as well as late-stage hadronic interactions, making it a potentially valuable observable to probe both the partonic and hadronic phases~\cite{heinz:2010,s_bass:1998,Sahoo:2023sgk}.
Since the reaction plane cannot be determined experimentally, the directed flow coefficient is commonly measured with respect to the event plane. The measured $v_{1}^{EP}$ using event plane method can be decomposed into rapidity-odd and rapidity-even components. The rapidity-even component may arise from different sources, such as event-by-event early-stage fluctuations, two-particle scattering and recoil, dijets, color-string (LUND) fragmentation, and momentum conservation.
The rapidity-odd component $v_{1}^{\mathrm{odd}}$, the focus of this paper, may result at these collision energies from a repulsive sideward deflection of the particles
and is antisymmetric about mid-rapidity. The sideward motion encoded in $v_{1}^{\mathrm{odd}}$ develops during the initial evolution of the strongly interacting medium and can later be modified by rescattering among hadrons, thereby carrying information from multiple stages of the collision evolution.

\begin{figure}
\includegraphics[scale=0.4]{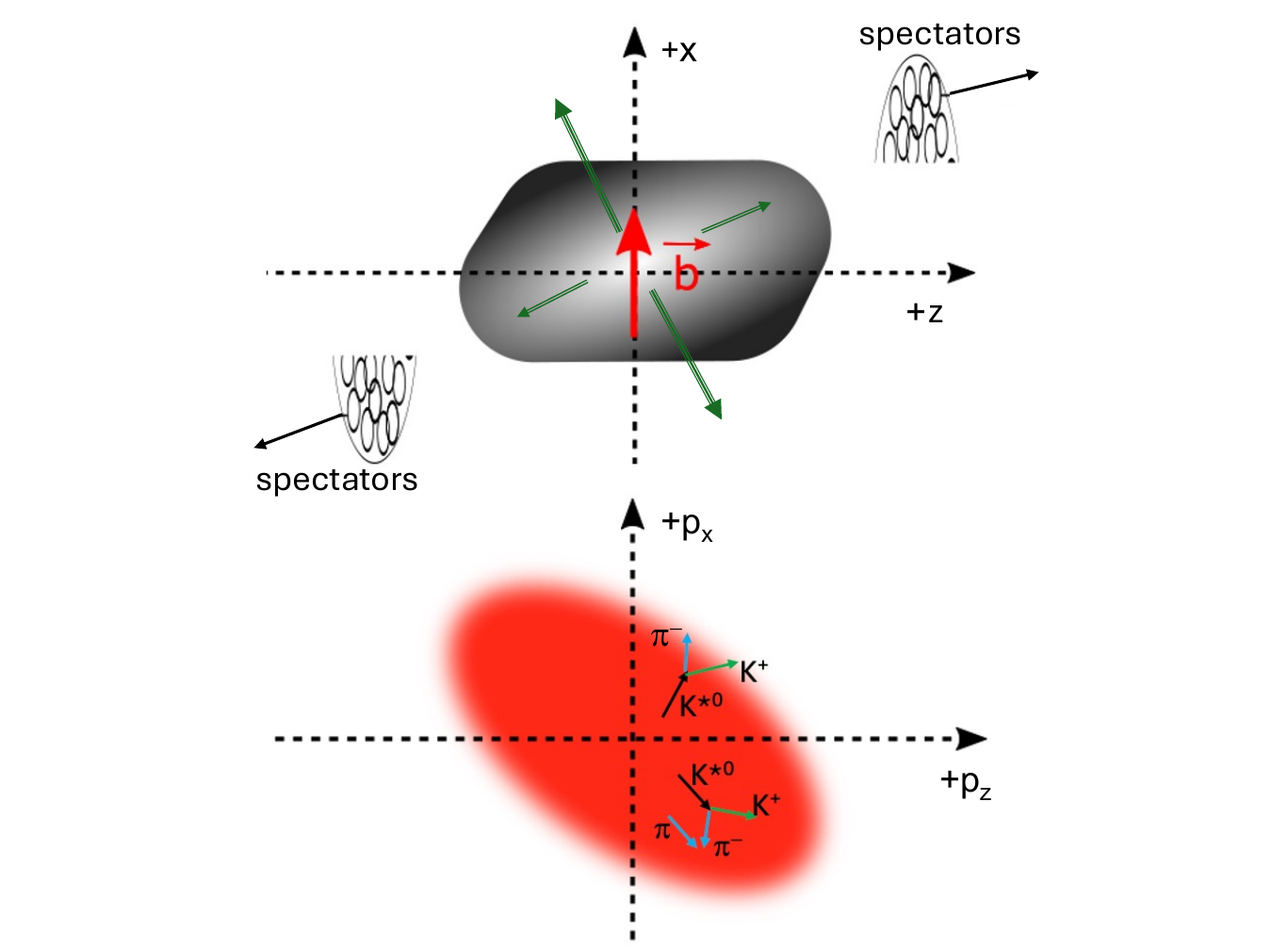}
\caption{The figure defines the event coordinate system used in the STAR experiment, where the beam direction defines the longitudinal $z$-axis. The positive $x$-direction (or impact parameter $\vec{b}$) is fixed 
by the deflection of spectators moving along the positive $z$-direction. The directions of the initial 
pressure gradient are shown by green arrows in the $x$-$z$ plane. 
Anisotropic flow of produced particles in momentum space is shown by the lower, red-shaded region in the 
$p_{x}$-$p_{z}$ plane. 
Particles emitted towards the direction of the impact parameter ($\vec{b}$) correspond to 
$v_{1}^{\mathrm{odd}}>0$, while emission opposite to the impact parameter corresponds to 
$v_{1}^{\mathrm{odd}}<0$. The re-scattering of a pion from the $K^{*0}$ decay with another pion is illustrated in the region $(p_{x}< 0,\ p_{z} > 0)$. 
Picture courtesy: T. Parida et al.~\cite{Parida:2023tdx}.}
\label{fig0}
\end{figure}

Figure~\ref{fig0} presents a schematic picture of the initial overlap geometry (grey shaded region) and 
the flow of particles in momentum coordinates (red shaded region) in a non-central heavy-ion collision. 
The beam direction is along the $z$-axis. The positive $x$-direction (or impact parameter $\vec{b}$) is 
determined by the deflection of spectators moving along the positive $z$-direction, which is outward at these higher collision energies~\cite{nida_2026}. Particles emitted in the 
direction of the impact parameter correspond to $v_{1}^{\mathrm{odd}}>0$, while emission opposite the 
impact parameter corresponds to $v_{1}^{\mathrm{odd}}<0$. In this picture, re-scattering effects on the $K^{*0}$ are intrinsically phase-space dependent as the $K^{*0}$ mostly decays within 
an anisotropic hadronic medium due to its very short lifetime. As illustrated in Fig.~\ref{fig0}, the regions 
($p_{x}<0$, $p_{z} >0$) and  ($p_{x}>0$, $p_{z} <0$) corresponds to a denser hadronic medium 
leading to enhanced re-scattering of the $K^{*0}$ decay daughters. In contrast, $K^{*0}$ mesons 
emitted toward  ($p_{x}>0$, $p_{z} >0$) and ($p_{x}<0$, $p_{z} <0$) are expected to encounter 
low density and shorter path lengths through the hadronic phase, resulting in a higher reconstruction 
probability. The resulting asymmetric reconstruction probability may distort the measured 
$v_{1}^{\mathrm{odd}}$ relative to 
the directed flow generated during earlier stages of the collision.

A recent hydrodynamic study, based on a $(3+1)$-dimensional hybrid framework with a hadronic evolution stage modeled using Ultra-relativistic Quantum Molecular Dynamics (UrQMD), investigated the $v_{1}^{\mathrm{odd}}$ of $K^{*0}$, $\phi$, and charged kaons~\cite{Parida:2023tdx}. 
The rapidity ($y$) differential $v_{1}^{\mathrm{odd}}$ 
slope ($dv_{1}^{\mathrm{odd}}/dy$) differences between the $\phi$ and $K^{+}$ mesons remain consistently negative for 
all centrality classes. In contrast, the $v_{1}^{\mathrm{odd}}$ difference between the $K^{*0}$ and $K^{+}$ mesons 
exhibits a clear centrality dependence; it is negative when the $K^{*0}$ mesons reconstruction is not affected by hadronic interactions, but changes sign once interactions in the hadronic phase are included.  This reversal in slope is 
attributed to the asymmetric modification of the $K^{*0}$ yield with respect to the first-order event 
plane caused by late-stage hadronic interactions. Although the production yield of $K^{*0}$ 
has been extensively measured in heavy-ion collisions to investigate hadronic rescattering 
effects~\cite{STAR:2002npn,STAR:2022sir,STAR:2026kqj,ALICE:2021ptz,Sahoo:2023rko}, its directed flow provides a 
more sensitive observable for exploring the scattering dynamics and dissipative properties of the 
late-stage hadronic medium. 

In this Letter, we present the first measurement of rapidity-odd directed flow of $K^{*0}$ and 
$\overline{K^{*0}}$ mesons in Au+Au collisions at $\sqrt{s_{\mathrm{NN}}} = 14.6$, 19.6, and 27~GeV, using 
minimum-bias (MB) data collected by the STAR experiment during the second phase of the 
Beam Energy Scan (BES-II) program at the Relativistic Heavy Ion Collider (RHIC). 
Measurements of directed flow of charged kaons and $\phi$ mesons are also presented in Au+Au collisions at $\sqrt{s_{\mathrm{NN}}} = 14.6$, 19.6, and 27~GeV.
This analysis utilizes approximately 500, 900, and 400 million MB events at 14.6, 19.6, and 27~GeV, 
respectively.

\section{Experiment and methods}

The $K^{*0}$ and $\overline{K^{*0}}$ mesons are reconstructed by way of their hadronic decay channels, 
$K^{*0}~(\overline{K^{*0}}) \rightarrow K^{\pm} \pi^{\mp}$. Unless explicitly stated, $K^{*0}$ in the 
remainder of this paper denotes the combination of particle and antiparticle. Simillarly the $\phi$ meson is reconstructed via it's hadronic decay channel $\phi \rightarrow K^{+}K^{-}$ (branching ratio $49.2\%$). Charged-particle 
trajectories (tracks) are reconstructed using the Time Projection Chamber (TPC)~\cite{STAR:tpc} in the presence of a uniform magnetic field of 0.5 T. Collision centrality is determined by comparing the 
measured charged-particle multiplicity distribution within pseudo-rapidity $|\eta| < 0.5$ 
with a Glauber Monte Carlo simulation. Accepted charged-particle tracks are required to have at least 
15 space points within the TPC and a distance of closest approach (DCA) to the primary collision vertex 
less than 2 cm. The minimum TPC hit requirement helps to avoid the short tracks and the cut on the DCA selection reduces track contamination from secondary vertices.
For $K^{*0}$ reconstruction, only charged tracks within the pseudorapidity range $|\eta| < 1.0$ are considered.
Charged tracks are also required have transverse momentum 
$0.15 < p_{\mathrm{T}} < 10.0$~GeV/$c$. The longitudinal position of the primary vertex ($V_{z}$), measured 
along the beam line, is required to lie within $\pm$140~cm of the TPC fiducial center at 14.6 and 
19.6~GeV, and within $\pm$70~cm at 27~GeV. The different selections correspond to the different detector configurations and acceptance conditions during the data-taking periods.

The decay daughters of $K^{*0}$ and $\phi$ mesons are identified using the truncated mean of the 
ionization energy loss ($\mathrm{d}E/\mathrm{d}x$) measured in the TPC, along with the inverse velocity ($1/\beta$) 
information from the Time-of-Flight (TOF) detector~\cite{STAR:tof}. To identify pions and kaons, 
the measured $dE/dx$ is required to lie within two standard deviations of the expected values calculated 
using the Bethe–Bloch parameterization~\cite{Bichsel:2006cs}. For tracks matched to TOF hits, pions and kaons are further 
selected by requiring $|1/\beta - 1/\beta_{\mathrm{exp}}| < 0.04$, where $1/\beta_{\mathrm{exp}}$ 
is the expected inverse velocity based on the expected particle mass.

The rapidity-odd directed flow is calculated using the event 
plane method where
\begin{equation}
    v_{1}^{\mathrm{odd}} =\frac{\langle \rm{cos}\ (\phi-\Psi^{\rm EP}_{1}) \rangle}{R_{1}}|_{y-odd}.
\end{equation}
The first-order event plane angle ($\Psi^{\rm EP}_{1}$) is reconstructed using the east and west sections of the 
Event Plane Detector (EPD)~\cite{STAR:epd}, which covers the pseudorapidity range $2.1 < |\eta| < 5.1$ 
and the full 2$\pi$ azimuth. The angle of the East-side event plane is rotated by $\pi$ before combination with the West-side event plane to account for the opposite spectator deflections on the two sides of the beam. This makes the measured directed flow primarily sensitive to $v_{1}^{\mathrm{odd}}$, while suppressing rapidity-even contributions. A detailed description of the $v_{1}^{\mathrm{odd}}$ measurement with the EPD is provided in another publication by the STAR Collaboration~\cite{v1_epd_star}.
At lower beam energies, the relatively strong $v_{1}^{\mathrm{odd}}$ signal in this pseudorapidity region, enhances the event plane resolution ($R_{1}$). Furthermore, the large pseudorapidity 
gap ($\Delta \eta \geq$ 1.1) between the EPD and the TPC, significantly suppresses non-flow 
contributions.Event plane resolution is calculated by correlating event planes measured separately by 
the east and west sections of the EPD~\cite{art_pos}.  For mid-central collisions (10--40\%), the typical, full 
$\Psi^{\rm EP}_{1}$ resolution using the EPD ranges from 50\% to 70\%~\cite{v1_epd_star}.

The $K^{*0}$ ($\phi$) signal is reconstructed from the invariant-mass distribution of oppositely charged kaon--pion (kaon-kaon) pairs formed within the same event.  This distribution contains the $K^{*0}$  ($\phi$) signal peak superimposed on a large combinatorial background arising from uncorrelated $K^{\pm}\pi^{\mp}$ ($K^{+}K^{-}$) pairs. The combinatorial background is estimated using a 
pair-rotation method, where the momentum vectors of one of the decay daughter particle species are 
rotated 180$^\circ$ in the transverse plane, thus eliminating correlations. The signal yield ($s$) is 
extracted by subtracting this background yield ($b$) from the same-event invariant mass distribution. The detailed procedure about $K^{*0}$ and $\phi$ reconstruction can be found in ref.~\cite{STAR:2026kqj}.
The combined ($s+b$) directed flow as a function of $M_{\mathrm{inv}}$ is given by~\cite{vn_inv_mass}
\[
v_{1}^{s+b}(M_{\mathrm{inv}}) = v_{1}^{s} \frac{s}{s + b}(M_{\mathrm{inv}}) + v_{1}^{b} \frac{b}{s + b}(M_{\mathrm{inv}}),
\]
where $v_{1}^{s}$ represents the $v_{1}^{\mathrm{odd}}$ of the signal, and $v_{1}^{b}$ is the 
background component, modeled as a $1^{st}$ order polynomial function of invariant mass. The $v_{1}^{\mathrm{odd}}$ 
measurement was cross-checked using an alternative mixed-event background method, in which a daughter 
particle from one event is combined with a daughter particle from a different event having similar 
characteristics (e.g.,  similar centrality and vertex position). In addition, a conventional event-plane 
($\Psi^{\rm EP}_1$) method was employed, using the resonance yield in $\phi - \Psi^{\rm EP}_{1}$ bins, 
as a further cross-check~\cite{art_pos}. Both methods yield results consistent with the default method.

The systematic uncertainties in $v_{1}^{\mathrm{odd}}$ are evaluated by varying analysis selection 
criteria. The varied parameters are changes in the invariant mass fitting range, the residual background 
parameterizations, event and track quality cuts, and particle identification selection cuts. 
The final systematic uncertainty is estimated by the quadratic sum of the individual contributions. 
The magnitude of the systematic uncertainties varies with particle species and depends on collision 
energy, centrality, and transverse momentum. For mid-central collisions, the systematic uncertainty in 
the average directed flow $\langle v_{1}^{\mathrm{odd}} \rangle$ is between 1-5\% for charged kaons, 
between 10–20\% for $\phi$ mesons, and within 20–30\% for $K^{*0}$ mesons. In the case of $K^{*0}$ mesons, yield extraction constitutes the dominant source of systematic uncertainty, especially at low $p_{T}$ and for central collision events.

\begin{figure*}
\includegraphics[scale=0.8]{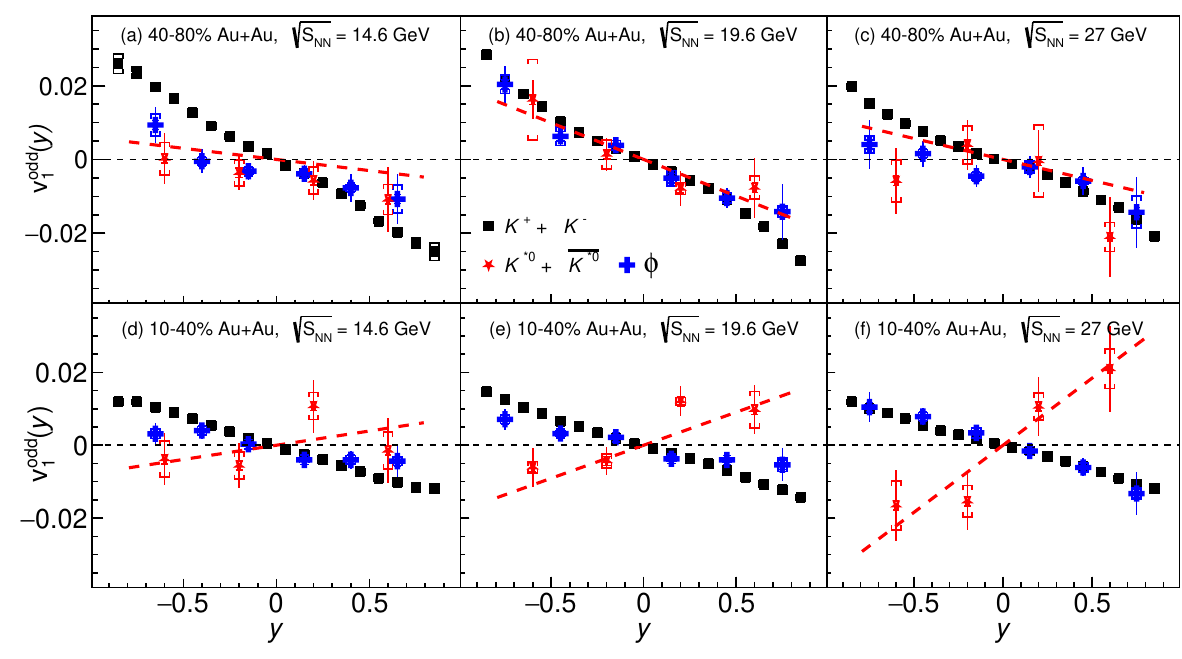}
\caption{Panels (a), (b), and (c) present the directed flow ($v_{1}^{\mathrm{odd}}(y)$), 
integrated over $p_{\mathrm{T}}$ within 0 $<$ $p_{\mathrm{T}}$ $<$ 5.0 GeV/$c$, versus rapidity (y) for $K^{*0}$ 
(red stars), charged kaons (black squares) and $\phi$ mesons (blue cross) for 40--80\% central Au+Au collisions 
at $\sqrt{s_{\mathrm{NN}}}$ = 14.6, 19.6 and 27 GeV, respectively. Panels (d), (e), and (f) present the same 
for 10-40\% central Au+Au collisions. The red dashed lines demonstrate the linear fit, 
$v_{1}^{\mathrm{odd}}(y) = p_{0} y$, for $K^{*0}$ mesons. Statistical and systematic uncertainties are added in quadrature while fitting the data points using $v_{1}^{\mathrm{odd}}(y) = p_{0} y$. }
\label{fig1}
\end{figure*}

\section{Results}
 Figure~\ref{fig1} shows the yield-weighted integral of directed flow $v_{1}^{\mathrm{odd}}$ over 
the transverse momentum ($p_{\mathrm{T}}$) interval 0 $<$ $p_{\mathrm{T}}$ $<$ 5.0 GeV/$c$, as a function of rapidity ($y$) 
for $K^{*0}$ (red stars), charged kaons (black squares), and $\phi$ mesons 
(blue cross) from Au+Au collisions at $\sqrt{s_{\mathrm{NN}}}$ = 14.6, 19.6, and 27~GeV. The top row 
corresponds to peripheral collisions (40--80\%), while the bottom row displays results for 
mid-central (10--40\%) collisions. The average number of participating nucleons $\langle N_{\mathrm{part}} \rangle$ in 
Au+Au collisions at $\sqrt{s_{\mathrm{NN}}} = 14.6$, 19.6, and 27~GeV are listed in table~\ref{tab:npart_values}.  As current statistics are sufficient only for a simple straight line fit, the slope ($dv_{1}^{\mathrm{odd}}/dy$) is extracted by 
fitting the data with a linear function constrained to pass through the origin, 
$v_{1}^{\mathrm{odd}}(y) = p_{0} y$. The red dashed lines display the linear fits for the $K^{*0}$ mesons. 
For peripheral collisions, the $v_{1}^{\mathrm{odd}}$ slope is negative for all particle species studied.
In contrast, for mid-central collisions the $K^{*0}$ meson $v_{1}^{\mathrm{odd}}$ displays a positive 
slope while the $\phi$ meson and charged kaons continue to exhibit negative slopes. It is important to 
note that the $\phi$ meson has a relatively long lifetime ($\sim$42 fm/$c$) that is much greater than that 
of the QGP and hadronic phases produced in heavy-ion collisions (of order $\sim$10 fm/$c$). In contrast, the $K^{*0}$ 
meson, with a shorter lifetime ($\sim$4 fm/$c$), decays before the medium reaches its final kinetic 
freeze-out stage. Consequently, the observed change in the $v_{1}^{\mathrm{odd}}$ slope of $K^{*0}$ mesons
from peripheral to mid-central collisions may be attributed to hadronic interaction effects in the 
late-stage hadronic medium, as predicted by model calculations~\cite{Parida:2023tdx}.

\begin{table}[htbp]
\centering
\caption{The average number of participating nucleons $\langle N_{\mathrm{part}} \rangle$ in 
Au+Au collisions at $\sqrt{s_{\mathrm{NN}}} = 14.6$, 19.6, and 27~GeV. }
\label{tab:npart_values}
\begin{tabular}{ccc}
\hline
$\sqrt{s_{NN}}$ (GeV) & 10--40\% & 40--80\% \\
\hline
27.0 & 166 $\pm$ 9 & 40 $\pm$ 8 \\
19.6 & 165 $\pm$ 10 & 40 $\pm$ 8 \\
14.6 & 164 $\pm$ 10 & 40 $\pm$ 8 \\
\hline
\end{tabular}
\end{table}

\begin{figure*}
\includegraphics[scale=0.8]{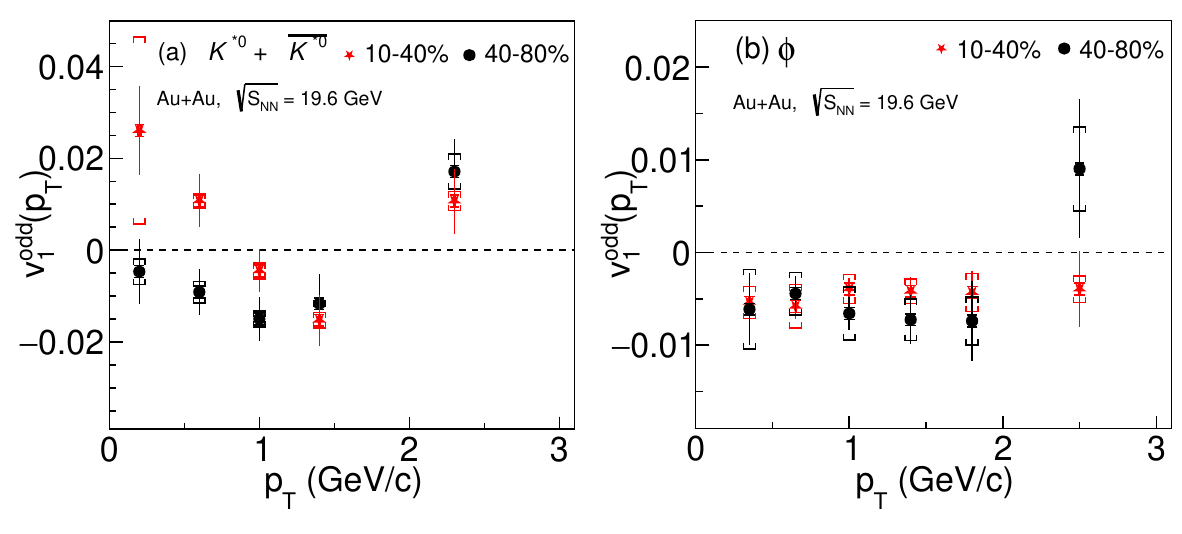}
\caption{The yield-weighted, rapidity-integrated $v_{1}^{\mathrm{odd}}(p_{\mathrm{T}})$ as a function of 
transverse momentum ($p_{\mathrm{T}}$) for 10–40\% and 40–80\% central Au+Au collisions at $\sqrt{s_{\mathrm{NN}}} = 19.6$~GeV 
is shown for $K^{*0}$ mesons in panel (a) and for $\phi$ mesons in panel (b).}
\label{fig2}
\end{figure*}

The odd component of $v_{1}$ is antisymmetric in rapidity, i.e., $v_{1}^{\mathrm{odd}}(-y, p_{\mathrm{T}}) = -v_{1}^{\mathrm{odd}}(y, p_{\mathrm{T}})$. As a result, it cancels out when integrated over a symmetric rapidity range. Therefore, to obtain the rapidity-integrated $v_{1}^{\mathrm{odd}}(p_{\mathrm{T}})$, the values at negative rapidity are multiplied by $-1$ before averaging. Figure~\ref{fig2}(a) presents the yield-weighted, rapidity-integrated 
$v_{1}^{\mathrm{odd}}$ calculated within $|y| < 1.0$ as a function of transverse momentum for 
$K^{*0}$ mesons. Similar results are shown in panel (b) for $\phi$ mesons.  Both panels include results
for the 10--40\% and 40--80\% centralities in Au+Au collisions at $\sqrt{s_{\mathrm{NN}}} = 19.6$~GeV. Results for $\sqrt{s_{\mathrm{NN}}}$ = 14.6 and 27 GeV are not shown because of limited statistics.
For $K^{*0}$ mesons, $v_{1}^{\mathrm{odd}}$ becomes positive at low $p_{\mathrm{T}} <$~1~GeV/$c$ in 
mid-central (10--40\%) collisions, whereas it remains negative in peripheral (40--80\%) collisions. 
In contrast, $v_{1}^{\mathrm{odd}}$ for $\phi$ mesons remains negative for both centralities. 
At higher transverse momentum ($p_{\mathrm{T}} >$~2~GeV/$c$) both $K^{*0}$ and $\phi$ mesons exhibit positive 
$v_{1}^{\mathrm{odd}}$ in (40--80\%) peripheral collisions, consistent with previous observations 
for inclusive charged-particle $v_{1}^{\mathrm{odd}}$~\cite{STAR:2019vcp} at these energies. The observation of a 
positive $v_{1}^{\mathrm{odd}}$ for $K^{*0}$ mesons at low $p_{\mathrm{T}}$ in mid-central collisions, 
in contrast with peripheral collisions, further supports the hypothesis that the sign change in 
$K^{*0}$ $dv_{1}^{\mathrm{odd}}/dy$ (as shown in Fig.~\ref{fig1}) may result from hadronic interactions suffered by the decay daughters with other particles in the medium, particularly for decay daughters originating from low-$p_{T}$ $K^{*0}$ mesons in dense, more-central collisions, where such effects are expected to be stronger~\cite{Sahoo:2023rko}.

\begin{figure*}
\includegraphics[scale=0.8]{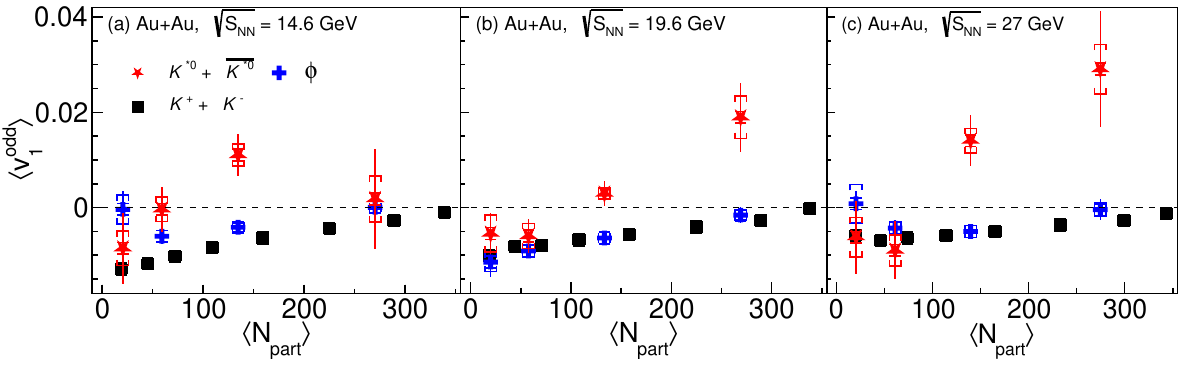}
\caption{Panels (a), (b), and (c) show the rapidity- and transverse-momentum integrated directed flow, 
$\langle v_{1}^{\mathrm{odd}} \rangle$, as a function of the average number of participating nucleons 
($\langle N_{\mathrm{part}} \rangle$) for $K^{*0}+\overline{K^{*0}}$, $K^{+}+K^{-}$, and $\phi$ mesons 
in Au+Au collisions at $\sqrt{s_{\mathrm{NN}}} = 14.6$, 19.6, and 27 GeV, respectively.}
\label{fig3}
\end{figure*}

Figure~\ref{fig3} presents the average $\langle v_{1}^{\mathrm{odd}} \rangle$ of $K^{*0}$ mesons as a 
function of $\langle N_{\mathrm{part}} \rangle$ in 
Au+Au collisions at $\sqrt{s_{\mathrm{NN}}} = 14.6$, 19.6, and 27~GeV. The rapidity- and $p_{\mathrm{T}}$-integrated 
directed flow $\langle v_{1}^{\mathrm{odd}} \rangle$ is a yield-weighted average of 
$v_{1}^{\mathrm{odd}} (p_{\mathrm{T}},y)$ within $|y|<1.0$ and 0 $<$ $p_{\mathrm{T}}$ $<$ 5.0 GeV/$c$ obtained after 
flipping the sign of $v_{1}^{\mathrm{odd}}$ for $y < 0$. The $\langle N_{\mathrm{part}} \rangle$ is
obtained from Monte Carlo (MC) Glauber simulations using a similar method as described 
in~\cite{glb_star,glb_kharzeev}.
The average $\langle v_{1}^{\mathrm{odd}}  \rangle$ of $K^{*0}$ are compared with those of 
$K^{+}+K^{-}$ and $\phi$ mesons. The $\langle v_{1}^{\mathrm{odd}}  \rangle$ of $K^{*0}$ exhibits a 
sign change from negative to positive as the collisions become more central. In contrast, 
$\langle v_{1}^{\mathrm{odd}} \rangle$ for both $\phi$ mesons and charged kaons remain negative.

\begin{figure}
\includegraphics[scale=0.4]{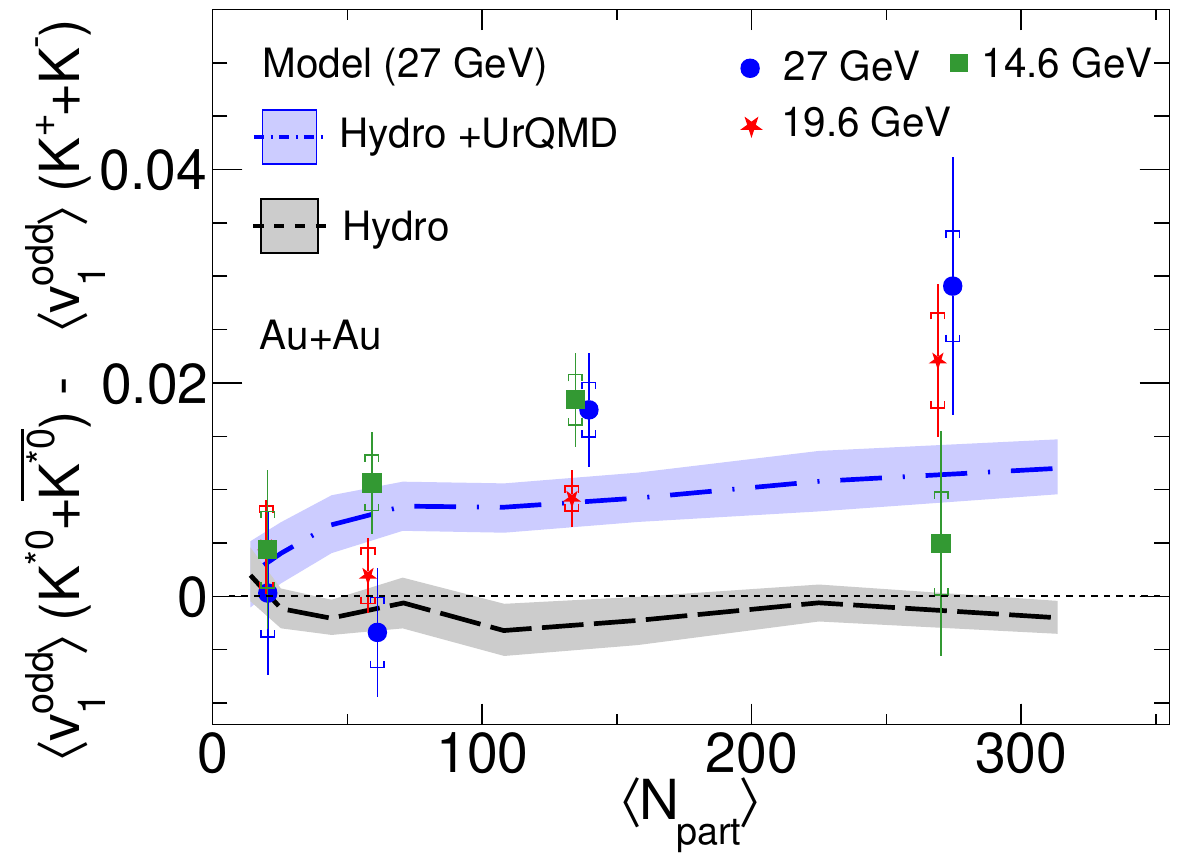}
\caption{The $\langle v_{1}^{\mathrm{odd}}  \rangle$ difference between $K^{*0}+\overline{K^{*0}}$ and 
$K^{+}+K^{-}$ as a function of $\langle N_{\mathrm{part}} \rangle$ in Au+Au collisions at 
$\sqrt{s_{\mathrm{NN}}} = 14.6$, 19.6, and 27 GeV. The results are compared with a hydrodynamical model, with 
and without a UrQMD afterburner, for the 27~GeV collisions. Bands on the model calculations represent statistical uncertainty.}
\label{fig4}
\end{figure}

Figure~\ref{fig4} presents the $\langle v_{1}^{\mathrm{odd}}  \rangle$ difference between 
$K^{*0}$ and $K^{+}+K^{-}$ as a function of $\langle N_{\mathrm{part}} \rangle$ in 
Au+Au collisions at $\sqrt{s_{\mathrm{NN}}} = 14.6$, 19.6, and 27~GeV. No obvious variation is found among the three measured collision energies. The values of $\langle v_{1}^{\mathrm{odd}}  \rangle$ differences between $K^{*0}$ and $K^{+}+K^{-}$ are shown in table~\ref{tab:v1_results} for 0--80\% centrality. 
For 0--80\%, their $\langle v_{1}^{\mathrm{odd}}  \rangle$ differences are consistent within 0.8 standard deviations in total uncertainty where the systematic uncertainties are assumed to be 100\% correlated. Weak energy dependence is also found by hydrodynamic model calculations~\cite{Parida:2023tdx}. The assumption of a 100\% correlation in the systematic uncertainty is considered  to be a good approximation since the contribution from uncorrelated systematic sources (e.g., detector effects, event selection variations)  is very small in this analysis.
Also in Fig.~\ref{fig4}, we compare our data to the prediction of the hydrodynamic model~\cite{Parida:2023tdx} at $\sqrt{s_{\mathrm{NN}}}$=27 GeV, both with and without a hadronic phase simulated using the UrQMD model~\cite{s_bass:1998}. While the $\langle v_{1}^{\mathrm{odd}}  \rangle$ 
difference is close to zero in the model without the hadronic phase, the inclusion of the hadronic 
afterburner produces a hint of weaker increase than that observed in data at the same energy, particularly in central and mid-central collisions. Integrated over the 0--80\% centrality range, our data support the scenario of a later-stage hadronic rescattering phase. The scenario of no hadronic scattering effects is excluded at the 99.4\% confidence level.
This may indicate that the centrality-dependent difference in directed flow 
between charged kaons and $K^{*0}$ resonances arises from a rapidity asymmetric suppression of the 
$K^{*0}$ yield with respect to the first-order event plane, that is likely to be driven by 
phase-space dependent hadronic interactions during the late stage of the medium's evolution, as 
predicted in~\cite{Parida:2023tdx}.

\begin{table}[htbp]
\centering
\caption{The $\langle v_{1}^{\mathrm{odd}}  \rangle$ differences between $K^{*0}$  and charged kaons in 0-80\% cenralities. }
\label{tab:v1_results}
\begin{tabular}{cccc}
\hline
$\sqrt{s_{NN}}$ (GeV) & $v_{1}(K^{*0})-v_{1}(K^{\pm})$ & Stat. Error & Syst. Error \\
\hline
27.0 & 0.00841 & 0.00340 & 0.00170 \\
19.6 & 0.00720 & 0.00186 & 0.00099 \\
14.6 & 0.01257 & 0.00285 & 0.00149 \\
\hline
\end{tabular}
\end{table}

\section{Summary}
We report measurements of directed flow ($v_{1}^{\mathrm{odd}}$) for charged kaons, $\phi$ 
mesons, and $K^{*0}$ resonances in Au+Au collisions at $\sqrt{s_{\mathrm{NN}}}=14.6$, 19.6, and 27 GeV. The data 
include the first $K^{*0}$ $v_{1}^{\mathrm{odd}}$ measurement in heavy-ion collisions. A clear, 
centrality-dependent difference is observed between the directed flow of charged kaons and $K^{*0}$ 
resonances that increases toward more-central collisions; the kaon–$\phi$ meson $v_{1}^{\mathrm{odd}}$
difference remains approximately centrality independent. 
We compare our measurements with hydrodynamic model predictions at a collision energy of $\sqrt{s_{\mathrm{NN}}}$ = 27 GeV, considering scenarios both with and without a hadronic phase implemented via the UrQMD model. In the absence of the hadronic phase, the $\langle v_{1}^{\mathrm{odd}} \rangle$ difference remains close to zero. Comparisons with hydrodynamic calculations incorporating a UrQMD-based hadronic afterburner suggest that the observed difference between the $v_{1}^{\mathrm{odd}}$ of charged kaons and $K^{*0}$ mesons is driven by hadronic rescattering effects. In particular, rescattering of the $K^{*0}$ decay daughters in the hadronic medium can lead to an asymmetric modification of the reconstructed $K^{*0}$ yield relative to the first-order event plane. When integrated over the 0–80\% centrality range,  the data exclude the configuration with no hadronic scattering at the 99.4\% confidence level. 
These results highlight the sensitivity of  short-lived resonances to hadronic re-scattering effects and challenge the interpretation of  $v_{1}^{\mathrm{odd}}$ in terms of partonic collectivity.  This work demonstrates that the directed flow of short-lived resonances  provides a sensitive probe of the hadronic phase and highlights the importance of incorporating  realistic hadronic scattering afterburners for reliable interpretation of collective flow observables.

\section{ACKNOWLEDGEMENTS}

We thank the RHIC Operations Group and SDCC at BNL, the NERSC Center at LBNL, and the Open Science Grid consortium for providing resources and support.  This work was supported in part by the Office of Nuclear Physics within the U.S. DOE Office of Science, the U.S. National Science Foundation, National Natural Science Foundation of China, Chinese Academy of Science, the Ministry of Science and Technology of China and the Chinese Ministry of Education, NSTC Taipei, the National Research Foundation of Korea, Czech Science Foundation and Ministry of Education, Youth and Sports of the Czech Republic, Hungarian National Research, Development and Innovation Office, New National Excellency Programme of the Hungarian Ministry of Human Capacities, Department of Atomic Energy and Department of Science and Technology of the Government of India, the National Science Centre and WUT ID-UB of Poland, German Bundes ministerium f\"ur Bildung, Wissenschaft, Forschung and Technologie (BMBF), Helmholtz Association, Ministry of Education, Culture, Sports, Science, and Technology (MEXT), Japan Society for the Promotion of Science (JSPS), and Agencia Nacional de Investigacion y Desarrollo de Chile (ANID), Chile.

\normalsize
\bibliographystyle{apsrev4-1}
\bibliography{reference}

\begin{thebibliography}{37}%
\makeatletter
\providecommand \@ifxundefined [1]{%
 \@ifx{#1\undefined}
}%
\providecommand \@ifnum [1]{%
 \ifnum #1\expandafter \@firstoftwo
 \else \expandafter \@secondoftwo
 \fi
}%
\providecommand \@ifx [1]{%
 \ifx #1\expandafter \@firstoftwo
 \else \expandafter \@secondoftwo
 \fi
}%
\providecommand \natexlab [1]{#1}%
\providecommand \enquote  [1]{``#1''}%
\providecommand \bibnamefont  [1]{#1}%
\providecommand \bibfnamefont [1]{#1}%
\providecommand \citenamefont [1]{#1}%
\providecommand \href@noop [0]{\@secondoftwo}%
\providecommand \href [0]{\begingroup \@sanitize@url \@href}%
\providecommand \@href[1]{\@@startlink{#1}\@@href}%
\providecommand \@@href[1]{\endgroup#1\@@endlink}%
\providecommand \@sanitize@url [0]{\catcode `\\12\catcode `\$12\catcode
  `\&12\catcode `\#12\catcode `\^12\catcode `\_12\catcode `\%12\relax}%
\providecommand \@@startlink[1]{}%
\providecommand \@@endlink[0]{}%
\providecommand \url  [0]{\begingroup\@sanitize@url \@url }%
\providecommand \@url [1]{\endgroup\@href {#1}{\urlprefix }}%
\providecommand \urlprefix  [0]{URL }%
\providecommand \Eprint [0]{\href }%
\providecommand \doibase [0]{http://dx.doi.org/}%
\providecommand \selectlanguage [0]{\@gobble}%
\providecommand \bibinfo  [0]{\@secondoftwo}%
\providecommand \bibfield  [0]{\@secondoftwo}%
\providecommand \translation [1]{[#1]}%
\providecommand \BibitemOpen [0]{}%
\providecommand \bibitemStop [0]{}%
\providecommand \bibitemNoStop [0]{.\EOS\space}%
\providecommand \EOS [0]{\spacefactor3000\relax}%
\providecommand \BibitemShut  [1]{\csname bibitem#1\endcsname}%
\let\auto@bib@innerbib\@empty
\bibitem [{\citenamefont {Adams}\ \emph {et~al.}(2005)\citenamefont {Adams}
  \emph {et~al.}}]{STAR:2005gfr}%
  \BibitemOpen
  \bibfield  {author} {\bibinfo {author} {\bibfnamefont {J.}~\bibnamefont
  {Adams}} \emph {et~al.} (\bibinfo {collaboration} {STAR}),\ }\href {\doibase
  10.1016/j.nuclphysa.2005.03.085} {\bibfield  {journal} {\bibinfo  {journal}
  {Nucl. Phys. A}\ }\textbf {\bibinfo {volume} {757}},\ \bibinfo {pages} {102}
  (\bibinfo {year} {2005})},\ \Eprint {http://arxiv.org/abs/nucl-ex/0501009}
  {arXiv:nucl-ex/0501009} \BibitemShut {NoStop}%
\bibitem [{\citenamefont {Becattini}\ and\ \citenamefont
  {Heinz}(1997)}]{Becattini:1997rv}%
  \BibitemOpen
  \bibfield  {author} {\bibinfo {author} {\bibfnamefont {F.}~\bibnamefont
  {Becattini}}\ and\ \bibinfo {author} {\bibfnamefont {U.~W.}\ \bibnamefont
  {Heinz}},\ }\href {\doibase 10.1007/s002880050551} {\bibfield  {journal}
  {\bibinfo  {journal} {Z. Phys. C}\ }\textbf {\bibinfo {volume} {76}},\
  \bibinfo {pages} {269} (\bibinfo {year} {1997})},\ \bibinfo {note} {[Erratum:
  Z.Phys.C 76, 578 (1997)]},\ \Eprint {http://arxiv.org/abs/hep-ph/9702274}
  {arXiv:hep-ph/9702274} \BibitemShut {NoStop}%
\bibitem [{\citenamefont {Braun-Munzinger}\ \emph {et~al.}(2001)\citenamefont
  {Braun-Munzinger}, \citenamefont {Magestro}, \citenamefont {Redlich},\ and\
  \citenamefont {Stachel}}]{Braun-Munzinger:2001hwo}%
  \BibitemOpen
  \bibfield  {author} {\bibinfo {author} {\bibfnamefont {P.}~\bibnamefont
  {Braun-Munzinger}}, \bibinfo {author} {\bibfnamefont {D.}~\bibnamefont
  {Magestro}}, \bibinfo {author} {\bibfnamefont {K.}~\bibnamefont {Redlich}}, \
  and\ \bibinfo {author} {\bibfnamefont {J.}~\bibnamefont {Stachel}},\ }\href
  {\doibase 10.1016/S0370-2693(01)01069-3} {\bibfield  {journal} {\bibinfo
  {journal} {Phys. Lett. B}\ }\textbf {\bibinfo {volume} {518}},\ \bibinfo
  {pages} {41} (\bibinfo {year} {2001})},\ \Eprint
  {http://arxiv.org/abs/hep-ph/0105229} {arXiv:hep-ph/0105229} \BibitemShut
  {NoStop}%
\bibitem [{\citenamefont {Xu}\ and\ \citenamefont {Kaneta}(2002)}]{Xu:2001zj}%
  \BibitemOpen
  \bibfield  {author} {\bibinfo {author} {\bibfnamefont {N.}~\bibnamefont
  {Xu}}\ and\ \bibinfo {author} {\bibfnamefont {M.}~\bibnamefont {Kaneta}},\
  }\href {\doibase 10.1016/S0375-9474(01)01377-X} {\bibfield  {journal}
  {\bibinfo  {journal} {Nucl. Phys. A}\ }\textbf {\bibinfo {volume} {698}},\
  \bibinfo {pages} {306} (\bibinfo {year} {2002})},\ \Eprint
  {http://arxiv.org/abs/nucl-ex/0104021} {arXiv:nucl-ex/0104021} \BibitemShut
  {NoStop}%
\bibitem [{\citenamefont {Brown}\ and\ \citenamefont
  {Rho}(1991)}]{Brown:1991kk}%
  \BibitemOpen
  \bibfield  {author} {\bibinfo {author} {\bibfnamefont {G.~E.}\ \bibnamefont
  {Brown}}\ and\ \bibinfo {author} {\bibfnamefont {M.}~\bibnamefont {Rho}},\
  }\href {\doibase 10.1103/PhysRevLett.66.2720} {\bibfield  {journal} {\bibinfo
   {journal} {Phys. Rev. Lett.}\ }\textbf {\bibinfo {volume} {66}},\ \bibinfo
  {pages} {2720} (\bibinfo {year} {1991})}\BibitemShut {NoStop}%
\bibitem [{\citenamefont {Markert}\ \emph {et~al.}(2008)\citenamefont
  {Markert}, \citenamefont {Bellwied},\ and\ \citenamefont
  {Vitev}}]{Markert:2008jc}%
  \BibitemOpen
  \bibfield  {author} {\bibinfo {author} {\bibfnamefont {C.}~\bibnamefont
  {Markert}}, \bibinfo {author} {\bibfnamefont {R.}~\bibnamefont {Bellwied}}, \
  and\ \bibinfo {author} {\bibfnamefont {I.}~\bibnamefont {Vitev}},\ }\href
  {\doibase 10.1016/j.physletb.2008.08.073} {\bibfield  {journal} {\bibinfo
  {journal} {Phys. Lett. B}\ }\textbf {\bibinfo {volume} {669}},\ \bibinfo
  {pages} {92} (\bibinfo {year} {2008})},\ \Eprint
  {http://arxiv.org/abs/0807.1509} {arXiv:0807.1509 [nucl-th]} \BibitemShut
  {NoStop}%
\bibitem [{\citenamefont
  {Schaffner-Bielich}(2000)}]{Schaffner-Bielich:1999cux}%
  \BibitemOpen
  \bibfield  {author} {\bibinfo {author} {\bibfnamefont {J.}~\bibnamefont
  {Schaffner-Bielich}},\ }\href {\doibase 10.1103/PhysRevLett.84.3261}
  {\bibfield  {journal} {\bibinfo  {journal} {Phys. Rev. Lett.}\ }\textbf
  {\bibinfo {volume} {84}},\ \bibinfo {pages} {3261} (\bibinfo {year}
  {2000})},\ \Eprint {http://arxiv.org/abs/hep-ph/9906361}
  {arXiv:hep-ph/9906361} \BibitemShut {NoStop}%
\bibitem [{\citenamefont {Rapp}\ and\ \citenamefont
  {Wambach}(2000)}]{Rapp:1999ej}%
  \BibitemOpen
  \bibfield  {author} {\bibinfo {author} {\bibfnamefont {R.}~\bibnamefont
  {Rapp}}\ and\ \bibinfo {author} {\bibfnamefont {J.}~\bibnamefont {Wambach}},\
  }\href {\doibase 10.1007/0-306-47101-9_1} {\bibfield  {journal} {\bibinfo
  {journal} {Adv. Nucl. Phys.}\ }\textbf {\bibinfo {volume} {25}},\ \bibinfo
  {pages} {1} (\bibinfo {year} {2000})},\ \Eprint
  {http://arxiv.org/abs/hep-ph/9909229} {arXiv:hep-ph/9909229} \BibitemShut
  {NoStop}%
\bibitem [{\citenamefont {Abelev}\ \emph {et~al.}(2006)\citenamefont {Abelev}
  \emph {et~al.}}]{STAR:2006vhb}%
  \BibitemOpen
  \bibfield  {author} {\bibinfo {author} {\bibfnamefont {B.~I.}\ \bibnamefont
  {Abelev}} \emph {et~al.} (\bibinfo {collaboration} {STAR}),\ }\href {\doibase
  10.1103/PhysRevLett.97.132301} {\bibfield  {journal} {\bibinfo  {journal}
  {Phys. Rev. Lett.}\ }\textbf {\bibinfo {volume} {97}},\ \bibinfo {pages}
  {132301} (\bibinfo {year} {2006})},\ \Eprint
  {http://arxiv.org/abs/nucl-ex/0604019} {arXiv:nucl-ex/0604019} \BibitemShut
  {NoStop}%
\bibitem [{\citenamefont {Rapp}\ and\ \citenamefont
  {Shuryak}(2001)}]{Rapp:2000gy}%
  \BibitemOpen
  \bibfield  {author} {\bibinfo {author} {\bibfnamefont {R.}~\bibnamefont
  {Rapp}}\ and\ \bibinfo {author} {\bibfnamefont {E.~V.}\ \bibnamefont
  {Shuryak}},\ }\href {\doibase 10.1103/PhysRevLett.86.2980} {\bibfield
  {journal} {\bibinfo  {journal} {Phys. Rev. Lett.}\ }\textbf {\bibinfo
  {volume} {86}},\ \bibinfo {pages} {2980} (\bibinfo {year} {2001})},\ \Eprint
  {http://arxiv.org/abs/hep-ph/0008326} {arXiv:hep-ph/0008326} \BibitemShut
  {NoStop}%
\bibitem [{\citenamefont {Navas}\ \emph {et~al.}(2024)\citenamefont {Navas}
  \emph {et~al.}}]{ParticleDataGroup:2024cfk}%
  \BibitemOpen
  \bibfield  {author} {\bibinfo {author} {\bibfnamefont {S.}~\bibnamefont
  {Navas}} \emph {et~al.} (\bibinfo {collaboration} {Particle Data Group}),\
  }\href {\doibase 10.1103/PhysRevD.110.030001} {\bibfield  {journal} {\bibinfo
   {journal} {Phys. Rev. D}\ }\textbf {\bibinfo {volume} {110}},\ \bibinfo
  {pages} {030001} (\bibinfo {year} {2024})}\BibitemShut {NoStop}%
\bibitem [{\citenamefont {Bleicher}\ and\ \citenamefont
  {Stoecker}(2004)}]{Bleicher:2003ij}%
  \BibitemOpen
  \bibfield  {author} {\bibinfo {author} {\bibfnamefont {M.}~\bibnamefont
  {Bleicher}}\ and\ \bibinfo {author} {\bibfnamefont {H.}~\bibnamefont
  {Stoecker}},\ }\href {\doibase 10.1088/0954-3899/30/1/010} {\bibfield
  {journal} {\bibinfo  {journal} {J. Phys. G}\ }\textbf {\bibinfo {volume}
  {30}},\ \bibinfo {pages} {S111} (\bibinfo {year} {2004})},\ \Eprint
  {http://arxiv.org/abs/hep-ph/0312278} {arXiv:hep-ph/0312278} \BibitemShut
  {NoStop}%
\bibitem [{\citenamefont {Knospe}\ \emph {et~al.}(2016)\citenamefont {Knospe},
  \citenamefont {Markert}, \citenamefont {Werner}, \citenamefont
  {Steinheimer},\ and\ \citenamefont {Bleicher}}]{Knospe:2015nva}%
  \BibitemOpen
  \bibfield  {author} {\bibinfo {author} {\bibfnamefont {A.~G.}\ \bibnamefont
  {Knospe}}, \bibinfo {author} {\bibfnamefont {C.}~\bibnamefont {Markert}},
  \bibinfo {author} {\bibfnamefont {K.}~\bibnamefont {Werner}}, \bibinfo
  {author} {\bibfnamefont {J.}~\bibnamefont {Steinheimer}}, \ and\ \bibinfo
  {author} {\bibfnamefont {M.}~\bibnamefont {Bleicher}},\ }\href {\doibase
  10.1103/PhysRevC.93.014911} {\bibfield  {journal} {\bibinfo  {journal} {Phys.
  Rev. C}\ }\textbf {\bibinfo {volume} {93}},\ \bibinfo {pages} {014911}
  (\bibinfo {year} {2016})},\ \Eprint {http://arxiv.org/abs/1509.07895}
  {arXiv:1509.07895 [nucl-th]} \BibitemShut {NoStop}%
\bibitem [{\citenamefont {Steinheimer}\ \emph {et~al.}(2017)\citenamefont
  {Steinheimer}, \citenamefont {Aichelin}, \citenamefont {Bleicher},\ and\
  \citenamefont {St\"ocker}}]{Steinheimer:2017vju}%
  \BibitemOpen
  \bibfield  {author} {\bibinfo {author} {\bibfnamefont {J.}~\bibnamefont
  {Steinheimer}}, \bibinfo {author} {\bibfnamefont {J.}~\bibnamefont
  {Aichelin}}, \bibinfo {author} {\bibfnamefont {M.}~\bibnamefont {Bleicher}},
  \ and\ \bibinfo {author} {\bibfnamefont {H.}~\bibnamefont {St\"ocker}},\
  }\href {\doibase 10.1103/PhysRevC.95.064902} {\bibfield  {journal} {\bibinfo
  {journal} {Phys. Rev. C}\ }\textbf {\bibinfo {volume} {95}},\ \bibinfo
  {pages} {064902} (\bibinfo {year} {2017})},\ \Eprint
  {http://arxiv.org/abs/1703.06638} {arXiv:1703.06638 [nucl-th]} \BibitemShut
  {NoStop}%
\bibitem [{\citenamefont {Bleicher}\ and\ \citenamefont
  {Aichelin}(2002)}]{Bleicher:2002dm}%
  \BibitemOpen
  \bibfield  {author} {\bibinfo {author} {\bibfnamefont {M.}~\bibnamefont
  {Bleicher}}\ and\ \bibinfo {author} {\bibfnamefont {J.}~\bibnamefont
  {Aichelin}},\ }\href {\doibase 10.1016/S0370-2693(02)01334-5} {\bibfield
  {journal} {\bibinfo  {journal} {Phys. Lett. B}\ }\textbf {\bibinfo {volume}
  {530}},\ \bibinfo {pages} {81} (\bibinfo {year} {2002})},\ \Eprint
  {http://arxiv.org/abs/hep-ph/0201123} {arXiv:hep-ph/0201123} \BibitemShut
  {NoStop}%
\bibitem [{\citenamefont {Adler}\ \emph {et~al.}(2002)\citenamefont {Adler}
  \emph {et~al.}}]{STAR:2002npn}%
  \BibitemOpen
  \bibfield  {author} {\bibinfo {author} {\bibfnamefont {C.}~\bibnamefont
  {Adler}} \emph {et~al.} (\bibinfo {collaboration} {STAR}),\ }\href {\doibase
  10.1103/PhysRevC.66.061901} {\bibfield  {journal} {\bibinfo  {journal} {Phys.
  Rev. C}\ }\textbf {\bibinfo {volume} {66}},\ \bibinfo {pages} {061901}
  (\bibinfo {year} {2002})},\ \Eprint {http://arxiv.org/abs/nucl-ex/0205015}
  {arXiv:nucl-ex/0205015} \BibitemShut {NoStop}%
\bibitem [{\citenamefont {Abdallah}\ \emph {et~al.}(2023)\citenamefont
  {Abdallah} \emph {et~al.}}]{STAR:2022sir}%
  \BibitemOpen
  \bibfield  {author} {\bibinfo {author} {\bibfnamefont {M.}~\bibnamefont
  {Abdallah}} \emph {et~al.} (\bibinfo {collaboration} {STAR}),\ }\href
  {\doibase 10.1103/PhysRevC.107.034907} {\bibfield  {journal} {\bibinfo
  {journal} {Phys. Rev. C}\ }\textbf {\bibinfo {volume} {107}},\ \bibinfo
  {pages} {034907} (\bibinfo {year} {2023})},\ \Eprint
  {http://arxiv.org/abs/2210.02909} {arXiv:2210.02909 [nucl-ex]} \BibitemShut
  {NoStop}%
\bibitem [{\citenamefont {Aboona}\ \emph {et~al.}(2026)\citenamefont {Aboona}
  \emph {et~al.}}]{STAR:2026kqj}%
  \BibitemOpen
  \bibfield  {author} {\bibinfo {author} {\bibfnamefont {B.~E.}\ \bibnamefont
  {Aboona}} \emph {et~al.} (\bibinfo {collaboration} {STAR}),\ }\href {\doibase
  10.1016/j.physletb.2026.140389} {\bibfield  {journal} {\bibinfo  {journal}
  {Phys. Lett. B}\ }\textbf {\bibinfo {volume} {876}},\ \bibinfo {pages}
  {140389} (\bibinfo {year} {2026})},\ \Eprint
  {http://arxiv.org/abs/2601.14884} {arXiv:2601.14884 [nucl-ex]} \BibitemShut
  {NoStop}%
\bibitem [{\citenamefont {Acharya}\ \emph {et~al.}(2022)\citenamefont {Acharya}
  \emph {et~al.}}]{ALICE:2021ptz}%
  \BibitemOpen
  \bibfield  {author} {\bibinfo {author} {\bibfnamefont {S.}~\bibnamefont
  {Acharya}} \emph {et~al.} (\bibinfo {collaboration} {ALICE}),\ }\href
  {\doibase 10.1103/PhysRevC.106.034907} {\bibfield  {journal} {\bibinfo
  {journal} {Phys. Rev. C}\ }\textbf {\bibinfo {volume} {106}},\ \bibinfo
  {pages} {034907} (\bibinfo {year} {2022})},\ \Eprint
  {http://arxiv.org/abs/2106.13113} {arXiv:2106.13113 [nucl-ex]} \BibitemShut
  {NoStop}%
\bibitem [{\citenamefont {Sahoo}\ \emph {et~al.}(2023)\citenamefont {Sahoo},
  \citenamefont {Nasim},\ and\ \citenamefont {Singha}}]{Sahoo:2023rko}%
  \BibitemOpen
  \bibfield  {author} {\bibinfo {author} {\bibfnamefont {A.~K.}\ \bibnamefont
  {Sahoo}}, \bibinfo {author} {\bibfnamefont {M.}~\bibnamefont {Nasim}}, \ and\
  \bibinfo {author} {\bibfnamefont {S.}~\bibnamefont {Singha}},\ }\href
  {\doibase 10.1103/PhysRevC.108.044904} {\bibfield  {journal} {\bibinfo
  {journal} {Phys. Rev. C}\ }\textbf {\bibinfo {volume} {108}},\ \bibinfo
  {pages} {044904} (\bibinfo {year} {2023})},\ \Eprint
  {http://arxiv.org/abs/2302.08177} {arXiv:2302.08177 [nucl-th]} \BibitemShut
  {NoStop}%
\bibitem [{\citenamefont {Sahoo}\ \emph {et~al.}(2025)\citenamefont {Sahoo},
  \citenamefont {Singha},\ and\ \citenamefont {Nasim}}]{Sahoo:2023dkv}%
  \BibitemOpen
  \bibfield  {author} {\bibinfo {author} {\bibfnamefont {A.~K.}\ \bibnamefont
  {Sahoo}}, \bibinfo {author} {\bibfnamefont {S.}~\bibnamefont {Singha}}, \
  and\ \bibinfo {author} {\bibfnamefont {M.}~\bibnamefont {Nasim}},\ }\href
  {\doibase 10.1088/1361-6471/ad8768} {\bibfield  {journal} {\bibinfo
  {journal} {J. Phys. G}\ }\textbf {\bibinfo {volume} {52}},\ \bibinfo {pages}
  {015101} (\bibinfo {year} {2025})},\ \Eprint
  {http://arxiv.org/abs/2307.06661} {arXiv:2307.06661 [nucl-th]} \BibitemShut
  {NoStop}%
\bibitem [{\citenamefont {Ollitrault}(1992)}]{Ollit:1992}%
  \BibitemOpen
  \bibfield  {author} {\bibinfo {author} {\bibfnamefont {J.-Y.}\ \bibnamefont
  {Ollitrault}},\ }\href {\doibase 10.1103/PhysRevD.46.229} {\bibfield
  {journal} {\bibinfo  {journal} {Phys. Rev. D}\ }\textbf {\bibinfo {volume}
  {46}},\ \bibinfo {pages} {229} (\bibinfo {year} {1992})}\BibitemShut
  {NoStop}%
\bibitem [{\citenamefont {Poskanzer}\ and\ \citenamefont
  {Voloshin}(1998)}]{art_pos}%
  \BibitemOpen
  \bibfield  {author} {\bibinfo {author} {\bibfnamefont {A.~M.}\ \bibnamefont
  {Poskanzer}}\ and\ \bibinfo {author} {\bibfnamefont {S.~A.}\ \bibnamefont
  {Voloshin}},\ }\href {\doibase 10.1103/PhysRevC.58.1671} {\bibfield
  {journal} {\bibinfo  {journal} {Phys. Rev. C}\ }\textbf {\bibinfo {volume}
  {58}},\ \bibinfo {pages} {1671} (\bibinfo {year} {1998})}\BibitemShut
  {NoStop}%
\bibitem [{\citenamefont {Heinz}()}]{heinz:2010}%
  \BibitemOpen
  \bibfield  {author} {\bibinfo {author} {\bibfnamefont {U.~W.}\ \bibnamefont
  {Heinz}},\ }\href {\doibase 10.48550/arXiv.0901.4355} {\
  10.48550/arXiv.0901.4355},\ \Eprint {http://arxiv.org/abs/0901.4355}
  {arXiv:0901.4355 [nucl-th]} \BibitemShut {NoStop}%
\bibitem [{\citenamefont {Bass}\ \emph {et~al.}(1998)\citenamefont {Bass} \emph
  {et~al.}}]{s_bass:1998}%
  \BibitemOpen
  \bibfield  {author} {\bibinfo {author} {\bibfnamefont {S.~A.}\ \bibnamefont
  {Bass}} \emph {et~al.},\ }\href {\doibase 10.1016/S0146-6410(98)00058-1}
  {\bibfield  {journal} {\bibinfo  {journal} {Prog. Part. Nucl. Phys.}\
  }\textbf {\bibinfo {volume} {41}},\ \bibinfo {pages} {225} (\bibinfo {year}
  {1998})}\BibitemShut {NoStop}%
\bibitem [{\citenamefont {Sahoo}\ \emph {et~al.}(2024)\citenamefont {Sahoo},
  \citenamefont {Dixit}, \citenamefont {Nasim},\ and\ \citenamefont
  {Singha}}]{Sahoo:2023sgk}%
  \BibitemOpen
  \bibfield  {author} {\bibinfo {author} {\bibfnamefont {A.~K.}\ \bibnamefont
  {Sahoo}}, \bibinfo {author} {\bibfnamefont {P.}~\bibnamefont {Dixit}},
  \bibinfo {author} {\bibfnamefont {M.}~\bibnamefont {Nasim}}, \ and\ \bibinfo
  {author} {\bibfnamefont {S.}~\bibnamefont {Singha}},\ }\href {\doibase
  10.1142/S0217732324500159} {\bibfield  {journal} {\bibinfo  {journal} {Mod.
  Phys. Lett. A}\ }\textbf {\bibinfo {volume} {39}},\ \bibinfo {pages}
  {2450015} (\bibinfo {year} {2024})},\ \Eprint
  {http://arxiv.org/abs/2309.14253} {arXiv:2309.14253 [nucl-th]} \BibitemShut
  {NoStop}%
\bibitem [{\citenamefont {Voloshin}\ and\ \citenamefont
  {Niida}(2016)}]{nida_2026}%
  \BibitemOpen
  \bibfield  {author} {\bibinfo {author} {\bibfnamefont {S.~A.}\ \bibnamefont
  {Voloshin}}\ and\ \bibinfo {author} {\bibfnamefont {T.}~\bibnamefont
  {Niida}},\ }\href {\doibase 10.1103/PhysRevC.94.021901} {\bibfield  {journal}
  {\bibinfo  {journal} {Phys. Rev. C}\ }\textbf {\bibinfo {volume} {94}},\
  \bibinfo {pages} {021901(R)} (\bibinfo {year} {2016})}\BibitemShut {NoStop}%
\bibitem [{\citenamefont {Parida}\ \emph {et~al.}(2024)\citenamefont {Parida},
  \citenamefont {Chatterjee},\ and\ \citenamefont {Nasim}}]{Parida:2023tdx}%
  \BibitemOpen
  \bibfield  {author} {\bibinfo {author} {\bibfnamefont {T.}~\bibnamefont
  {Parida}}, \bibinfo {author} {\bibfnamefont {S.}~\bibnamefont {Chatterjee}},
  \ and\ \bibinfo {author} {\bibfnamefont {M.}~\bibnamefont {Nasim}},\ }\href
  {\doibase 10.1103/PhysRevC.109.044905} {\bibfield  {journal} {\bibinfo
  {journal} {Phys. Rev. C}\ }\textbf {\bibinfo {volume} {109}},\ \bibinfo
  {pages} {044905} (\bibinfo {year} {2024})},\ \Eprint
  {http://arxiv.org/abs/2312.06359} {arXiv:2312.06359 [nucl-th]} \BibitemShut
  {NoStop}%
\bibitem [{\citenamefont {Ackermann}\ \emph {et~al.}(2003)\citenamefont
  {Ackermann} \emph {et~al.}}]{STAR:tpc}%
  \BibitemOpen
  \bibfield  {author} {\bibinfo {author} {\bibfnamefont {K.~H.}\ \bibnamefont
  {Ackermann}} \emph {et~al.} (\bibinfo {collaboration} {STAR}),\ }\href
  {\doibase 10.1016/S0168-9002(02)01960-5} {\bibfield  {journal} {\bibinfo
  {journal} {Nucl. Instrum. Meth. A}\ }\textbf {\bibinfo {volume} {499}},\
  \bibinfo {pages} {624} (\bibinfo {year} {2003})}\BibitemShut {NoStop}%
\bibitem [{\citenamefont {Llope}\ \emph {et~al.}(2004)\citenamefont {Llope}
  \emph {et~al.}}]{STAR:tof}%
  \BibitemOpen
  \bibfield  {author} {\bibinfo {author} {\bibfnamefont {W.~J.}\ \bibnamefont
  {Llope}} \emph {et~al.},\ }\href {\doibase 10.1016/j.nima.2003.11.414}
  {\bibfield  {journal} {\bibinfo  {journal} {Nucl. Instrum. Meth. A}\ }\textbf
  {\bibinfo {volume} {522}},\ \bibinfo {pages} {252} (\bibinfo {year}
  {2004})}\BibitemShut {NoStop}%
\bibitem [{\citenamefont {Bichsel}(2006)}]{Bichsel:2006cs}%
  \BibitemOpen
  \bibfield  {author} {\bibinfo {author} {\bibfnamefont {H.}~\bibnamefont
  {Bichsel}},\ }\href {\doibase 10.1016/j.nima.2006.03.009} {\bibfield
  {journal} {\bibinfo  {journal} {Nucl. Instrum. Meth. A}\ }\textbf {\bibinfo
  {volume} {562}},\ \bibinfo {pages} {154} (\bibinfo {year}
  {2006})}\BibitemShut {NoStop}%
\bibitem [{\citenamefont {Adams}\ \emph {et~al.}(2020)\citenamefont {Adams}
  \emph {et~al.}}]{STAR:epd}%
  \BibitemOpen
  \bibfield  {author} {\bibinfo {author} {\bibfnamefont {J.}~\bibnamefont
  {Adams}} \emph {et~al.},\ }\href {\doibase
  https://doi.org/10.1016/j.nima.2020.163970} {\bibfield  {journal} {\bibinfo
  {journal} {Nuclear Instruments and Methods in Physics Research Section A:
  Accelerators, Spectrometers, Detectors and Associated Equipment}\ }\textbf
  {\bibinfo {volume} {968}},\ \bibinfo {pages} {163970} (\bibinfo {year}
  {2020})},\ \Eprint {http://arxiv.org/abs/1912.05243} {arXiv:1912.05243
  [physics]} \BibitemShut {NoStop}%
\bibitem [{\citenamefont {Abdulhamid}\ \emph {et~al.}(2025)\citenamefont
  {Abdulhamid} \emph {et~al.}}]{v1_epd_star}%
  \BibitemOpen
  \bibfield  {author} {\bibinfo {author} {\bibfnamefont {M.~I.}\ \bibnamefont
  {Abdulhamid}} \emph {et~al.} (\bibinfo {collaboration} {STAR
  Collaboration}),\ }\href {\doibase 10.1103/PhysRevC.111.014906} {\bibfield
  {journal} {\bibinfo  {journal} {Phys. Rev. C}\ }\textbf {\bibinfo {volume}
  {111}},\ \bibinfo {pages} {014906} (\bibinfo {year} {2025})}\BibitemShut
  {NoStop}%
\bibitem [{\citenamefont {Borghini}\ and\ \citenamefont
  {Ollitrault}(2004)}]{vn_inv_mass}%
  \BibitemOpen
  \bibfield  {author} {\bibinfo {author} {\bibfnamefont {N.}~\bibnamefont
  {Borghini}}\ and\ \bibinfo {author} {\bibfnamefont {J.-Y.}\ \bibnamefont
  {Ollitrault}},\ }\href {\doibase 10.1103/PhysRevC.70.064905} {\bibfield
  {journal} {\bibinfo  {journal} {Phys. Rev. C}\ }\textbf {\bibinfo {volume}
  {70}},\ \bibinfo {pages} {064905} (\bibinfo {year} {2004})}\BibitemShut
  {NoStop}%
\bibitem [{\citenamefont {Adam}\ \emph {et~al.}(2020)\citenamefont {Adam} \emph
  {et~al.}}]{STAR:2019vcp}%
  \BibitemOpen
  \bibfield  {author} {\bibinfo {author} {\bibfnamefont {J.}~\bibnamefont
  {Adam}} \emph {et~al.} (\bibinfo {collaboration} {STAR}),\ }\href {\doibase
  10.1103/PhysRevC.101.024905} {\bibfield  {journal} {\bibinfo  {journal}
  {Phys. Rev. C}\ }\textbf {\bibinfo {volume} {101}},\ \bibinfo {pages}
  {024905} (\bibinfo {year} {2020})},\ \Eprint
  {http://arxiv.org/abs/1908.03585} {arXiv:1908.03585 [nucl-ex]} \BibitemShut
  {NoStop}%
\bibitem [{\citenamefont {Abelev}\ \emph {et~al.}(2010)\citenamefont {Abelev}
  \emph {et~al.}}]{glb_star}%
  \BibitemOpen
  \bibfield  {author} {\bibinfo {author} {\bibfnamefont {B.~I.}\ \bibnamefont
  {Abelev}} \emph {et~al.} (\bibinfo {collaboration} {STAR}),\ }\href {\doibase
  10.1103/PhysRevC.81.024911} {\bibfield  {journal} {\bibinfo  {journal} {Phys.
  Rev. C}\ }\textbf {\bibinfo {volume} {81}},\ \bibinfo {pages} {024911}
  (\bibinfo {year} {2010})}\BibitemShut {NoStop}%
\bibitem [{\citenamefont {Kharzeev}\ and\ \citenamefont
  {Nardi}(2001)}]{glb_kharzeev}%
  \BibitemOpen
  \bibfield  {author} {\bibinfo {author} {\bibfnamefont {D.}~\bibnamefont
  {Kharzeev}}\ and\ \bibinfo {author} {\bibfnamefont {M.}~\bibnamefont
  {Nardi}},\ }\href {\doibase 10.1016/S0370-2693(01)00457-9} {\bibfield
  {journal} {\bibinfo  {journal} {Phys. Lett. B}\ }\textbf {\bibinfo {volume}
  {507}},\ \bibinfo {pages} {121} (\bibinfo {year} {2001})}\BibitemShut
  {NoStop}%
\end{thebibliography}%

\end{document}